\documentclass[
    ,final            
  ]
  {aipproc}

\usepackage{amsmath}
\usepackage{amssymb}

\layoutstyle{8x11single}

\begin{document}

\def \G{\mathcal{G}}

\title{Time-dependent Green's functions approach to nuclear reactions}


\author{Arnau Rios}{address={National Superconducting Cyclotron Laboratory and 
Department of Physics and Astronomy,
Michigan State University, East Lansing, MI 48824-1321, USA }}

\author{Pawel Danielewicz}{address={National Superconducting Cyclotron Laboratory and
Department of Physics and Astronomy,
Michigan State University, East Lansing, MI 48824-1321, USA },
altaddress={Kavli Institute for Theoretical Physics,
University of California,
Santa Barbara, CA 93106-4030, USA}}

\begin{abstract}
Nonequilibrium Green's functions represent underutilized means of studying the time evolution of quantum many-body systems.  In view of a rising computer power, an effort is underway to apply the Green's functions formalism to the dynamics of central nuclear reactions.  As the first step, mean-field evolution for the density matrix for colliding slabs is studied in one dimension.  Strategy to extend the dynamics to correlations is described.
\end{abstract}

\maketitle


\section{Introduction}
Nuclear reactions represent active means of studying nuclei, nuclear interactions and nuclear medium.  Central reactions disturb nuclei more than do the peripheral ones and can lead to nuclear fusion, multifragmentation and dramatic compression of nuclear matter, depending on incident energy.  Due to the increasing many-body nature of the central reactions with higher energies and due to the corresponding population of a multitude of final states, the description of those reactions has relied on the time domain and has remained primarily semiclassical.  Because of their semiclassical nature, the descriptions for central reactions have then remained somewhat disconnected from the descriptions employed for nuclear structure, peripheral reactions or giant nuclear excitations.
The coarseness of the central-reaction approaches, associated to their semiclassical nature, is not easy to quantify and this has proven to be a difficulty in linking correctly the observables from the reactions to the properties of nuclear systems under the special conditions created in those reactions.

Historically, the only practical approach to central collisions, that can be considered quantal, has been the Time-Dependent Hartree-Fock (TDHF) method~\cite{tdhf} and its extensions \cite{tohyama87,lacroix04}. The starting point for TDHF is the mean-field description for nuclei, with individual wave-function components propagated in time according to Schr\"odinger-like single-particle equations of motion.
The validity of TDHF relies on the presumption that correlations play a negligible role in the dynamics. However, with the weakening of Pauli principle as the excitation energy increases in actual reactions, correlations can lead to a fast thermalization of the population of single-particle states as well as to enhanced stopping.
Such phenomena are completely absent in any TDHF-like description.
In view of the issues above, it would seem important to develop a quantal approach to central reaction dynamics including the effects of correlations beyond the mean-field. These correlations are also known to play a significant role in the initial state~\cite{dickhoff05}.

The time-dependent Green's functions formalism~\cite{kadanoff} is a natural candidate for the sought theory. On one hand, equilibrium Green's functions can describe the microscopic and macroscopic properties of stationary strongly interacting many-body system, accommodating the effects of correlations \cite{dickhoff05}.  On the other hand, an extension of the approach to the time-dependent domain, by means of the Kadanoff-Baym (KB) equations, can also incorporate the effect of correlations onto the time evolution \cite{kadanoff}.  To date, the application of the nonequilibrium formalism to nonuniform quantum many-body systems has been relatively scarce, but this can be in large part attributed to an involved numerical effort which should become of a lesser concern as the computational power grows.  Besides central reactions, the approach could be used to study the response of nuclei to external time-dependent fields~\cite{lacroix04}.

A very important feature of the Green's functions formalism is the fact that, within certain approximations for the self-energies, the evolution of the observables is conserving, \emph{i.e.} the main conservation laws (energy, momentum, angular momentum, etc) are preserved during the time evolution \cite{baym61,baym62}. In addition, the propagator formalism naturally takes into account the quantal properties of the problem. The non-local features of quantum mechanics are for instance reflected into the off-diagonal components of the Green's functions in real space representation. Little is known about the structure of these components and their importance in the time-evolution of quantum many-body systems.

\section{Kadanoff-Baym equations}

The KB equations generally describe the evolution, under rather liberal assumptions, of expectation values of the products of two single-particle annihilation and creation operators, at different time arguments, i.e.\ 2-point Wightman functions.  In the context of nonequilibrium theory, these are called the single-particle Green's functions.  At equal time-arguments, the expectation values yield the single-particle density matrix.  For realistic modeling of reactions, a~nonuniform three-dimensional (3D) system needs to be studied.  First, however, we shall restrict to the less involved one-dimensional (1D) case to get acquainted with the effect that correlations might induce on the time evolution of the system. Here, in addition, we shall assume in the following that the initial state of the system is uncorrelated, i.e.\ describable by mean-field theory. This approximation will be eventually relaxed, leading to a modification of the KB equations which we shall again not treat in here \cite{danielewicz84a}.

The initial state of the systems is specified in terms of an $A$-body density operator $\hat \rho_0$ at time $t_0$. The one-body Green's functions are then defined as expectation values, with respect to $\hat \rho_0$,
of products of Heisenberg-picture creation $\hat a^\dagger(x,t)$ and destruction $\hat a(x,t)$ operators:
\begin{align}
\G^<(x_1,t_1; x_2,t_2) &=
i \left< \hat a^\dagger(x_2,t_2) \hat a(x_1,t_1) \right> \, , \\
\G^>(x_1,t_1; x_2,t_2) &=
- i \left< \hat a(x_1,t_1) \hat a^\dagger(x_2,t_2)  \right>  \, .
\end{align}
Up to a factor, the time-diagonal Green's function $\G^<$ reduces to the one-body density matrix.  The KB equations, governing the evolution of Green's functions in their arguments,
\begin{align}
	\left\{ i \hbar \frac{\partial}{\partial t_1} + \frac{ \hbar^2}{2m} \frac{\partial^2}{\partial x_1^2}  \right\} \G^{\lessgtr} (\mathbf{11'})
	&= \int \!\! \textrm{d} \bar{\mathbf{r}}_1 \Sigma_{HF}( \mathbf{1} \bar{\mathbf{1}} ) \G^{\lessgtr} ( \bar{\mathbf{1}} \mathbf{1}' )
	 + \int_{t_0}^{t_1} \!\!\! \textrm{d} \bar{\mathbf{1}} \, \Sigma^{+}( \mathbf{1} \bar{\mathbf{1}} ) \G^{\lessgtr} ( \bar{\mathbf{1}} \mathbf{1}' )
	+ \int_{t_0}^{t_{1'}} \!\!\! \textrm{d} \bar{\mathbf{1}} \, \Sigma^{\lessgtr}( \mathbf{1} \bar{\mathbf{1}} ) \G^{-} ( \bar{\mathbf{1}} \mathbf{1}' ) \, ,
        \label{eq:kb1} \\
	\left\{- i  \hbar \frac{\partial}{\partial t_{1'}} +  \frac{ \hbar ^2}{2m} \frac{\partial^2}{\partial x_{1'}^2} \right\} \G^{\lessgtr} (\mathbf{11'})
	& = \int \!\! \textrm{d} \bar{\mathbf{r}}_1 \G^{\lessgtr} ( \mathbf{1} \bar{\mathbf{1}} ) \Sigma_{HF}( \bar{\mathbf{1}} \mathbf{1}')
	+ \int_{t_0}^{t_{1}} \!\!\! \textrm{d} \bar{\mathbf{1}} \, \G^{+} ( \mathbf{1} \bar{\mathbf{1}} ) \Sigma^{\lessgtr}( \bar{\mathbf{1}} \mathbf{1}' )
	 + \int_{t_0}^{t_{1'}} \!\!\! \textrm{d} \bar{\mathbf{1}} \, \G^{\lessgtr} (\mathbf{1}  \bar{\mathbf{1}} ) \Sigma^{-}( \bar{\mathbf{1}}  \mathbf{1}' ) \, ,
         \label{eq:kb1p}
\end{align}
follow from considerations of the equations of motion for creation and destruction operators. In the above, the simplified notation $\mathbf{1}=(x_1,t_1)$ has been introduced  and the retarded and advanced functions are defined according to:
\begin{align}
F^{\pm}(\mathbf{1},\mathbf{2}) = F^\delta (\mathbf{1},\mathbf{2})
\pm \Theta\left[ \pm (t_1-t_2)\right] \left[ F^>(\mathbf{1},\mathbf{2}) - F^<(\mathbf{1},\mathbf{2}) \right] \, ,
\end{align}
with $F^\delta$ standing for a possible singular contribution at $t_1=t_2$. A~generalized self-energy $\Sigma(\mathbf{1},\mathbf{2})$ may be introduced~\cite{danielewicz84a} which accounts for all the interaction effects of the system, including the instantaneous Hartree-Fock contribution,  $\Sigma_{HF}(\mathbf{1},\mathbf{2})$, as well as the $\Sigma^\lessgtr$ terms, generated by correlation effects that go beyond the mean-field.

The complex integro-differential KB equations have to be solved in a self-consistent way, since the self-energies depend on the Green's functions that are being solved for.  This self-consistency simultaneously guarantees a microscopic consistency of the theory. In particular, for certain many-body approximations to the self-energy (those of the conserving type), one can show that the time evolution induced by the self-consistent KB equations preserves the conservation laws obeyed by the system as a whole \cite{baym62,danielewicz84a}. These conservation laws can therefore be used as tests for the numerical implementation of the equations.

An attractive feature of the KB equations is their generality. The time evolution induced by Eqs.~\eqref{eq:kb1} and \eqref{eq:kb1p} can easily describe different types of correlations, given in terms of different approximations to the self-energy. In addition, the equations do not focus on any particular physical system and can therefore be used to study different many-body problems. So far, the KB equations have been used to study the time evolution of uniform nuclear matter \cite{danielewicz84b,kohler95} and the uniform electron gas \cite{kwong98}, as well as inhomogeneous atomic and molecular systems \cite{dahlen07}.

The KB equations account for the so-called memory effects. From Eq.~\eqref{eq:kb1}, one can easily see that that the Green's function $\G^<$ at times $t_1$ and $t_{1'}$ depends on the Green's functions and self-energies at all the previous times t, $t_1>t>t_0$ and $t_{1'}>t>t_0$, via the time integrals on the r.h.s\ of the equation. In consequence, to find a solution of the equations one must keep track of all the previous time-steps. A strategy to attack this problem, before addressed in~\cite{danielewicz84b} and \cite{kohler99}, is discussed further on.

Finally, it is important to note that the KB equations are quantal and, in particular, respect the non-local features of quantum mechanics. This is directly reflected in the fact that the evolution, even in the 1D case, depends on two space and two time variables, $(x_1,t_1)$ and $(x_{1'},t_{1'})$. This doubling of variables might be a disadvantage in terms of computational cost, especially compared to other time-dependent approaches which do not share this feature. We believe, however, that the increase in computational cost (if any) might be compensated by the advantages of describing the system in terms of the Wightman's functions.

\section{Mean-field approximation}

\begin{figure}
  \includegraphics[height=.4\textheight]{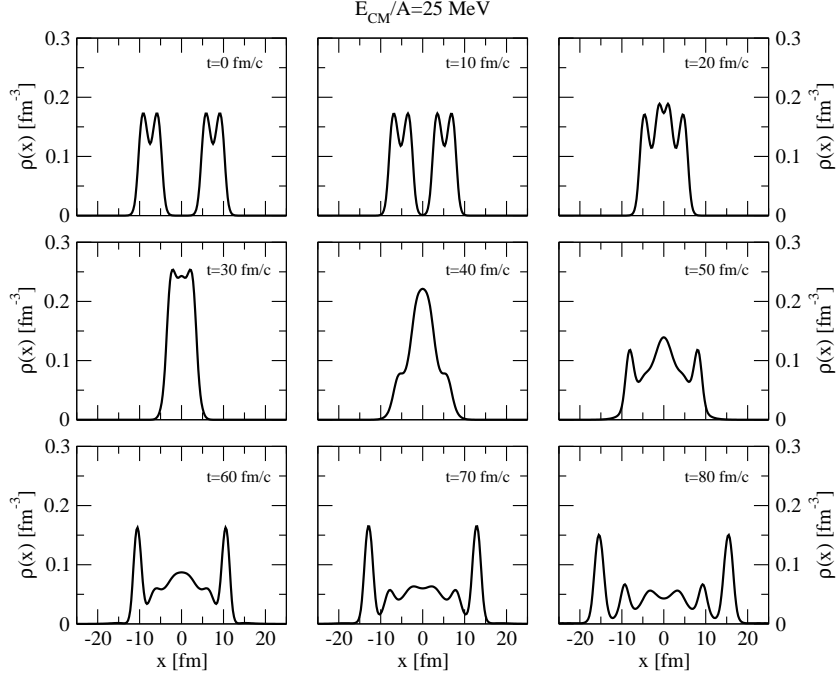}
  \caption{Time evolution of density for slabs colliding in the mean-field approximation.}
  \label{fig:denmat}
\end{figure}

The rather involved calculations associated with the KB equations simplify when one deals only with the Hartree-Fock part of the self-energy. The evolution of the correlation functions is then given in terms of the mean-field and there are no memory effects involved in the calculations. The KB equations reduce in that case to the TDHF equations for the evolution of single-particle density matrix for the system. This simple framework can be used as a starting point for the studies of a time-dependent approach formulated in terms of the density matrix, rather than in terms of single-particle wave-functions, as would have been standard for a TDHF approach.

Since this is meant as a first qualitative example, we have chosen a rather simple local density-dependent mean-field of the Skyrme-type:
\begin{align}
\Sigma_{HF}( \mathbf{1} \bar{\mathbf{1}} ) = \delta(t_1 - t_{1'}) \, \delta( x_1 -  x_{1'} )
\left[ \frac{3}{4} \, t_0 \, n(x_1,t) + \frac{2+\sigma}{16} \, t_3 \left[ n(x_1,t) \right]^{(\sigma+1)} \right] \, ,
\end{align}
with the parameters $t_0$, $t_3$ and $\sigma$ fitted to reproduce the saturation properties of infinite nuclear matter. The local density $n(x,t)$ corresponds actually to a 3D system and it is related to the 1D density by a scaling factor, $n(x,t) = -i \, \xi \, \G(x, t; x_{1'}=x, t_{1'}=t) $, where $\xi= (4 \, n^2_0 )^{1/3}$ and $n_0 = 0.16 \, \text{fm}^{-3}$ is the density of normal nuclear matter.  In the above, we assume a uniformity of the 3D system in the perpendicular $y$ and $z$ directions, as well as a spin-isospin degeneracy of $\nu=4$.

Since the Hartree-Fock interaction is local, one can implement the time evolution in a rather straightforward way by means of the so-called Split Operator Method (SOM) \cite{som}. This is based on the fact that, for infinitesimally short time-steps $\Delta t$, the evolution of the density matrix is formally given in terms of the single-particle kinetic $K$ and mean-field $U$ operators as:
\begin{align}
\G(x,t+\Delta t; x',t'+\Delta t') =
\text{e}^{ -i \frac{(\hat K+ \hat U) \Delta t}{\hbar} } \, \G(x, t ; x',t') \, \text{e}^{ i \frac{(\hat K + \hat U) \Delta t'}{\hbar} } \, .
\label{eq:SOM}
\end{align}
For short times, one can use the decomposition
$ \text{e}^{ i \frac{(\hat K+ \hat U) \Delta t}{\hbar}}
=  \text{e}^{ i \frac{ \hat U \Delta t}{2 \hbar}}
\text{e}^{ i \frac{\hat K \Delta t}{\hbar}}
\text{e}^{ i \frac{\hat U \Delta t}{2 \hbar}} + O(\Delta t^3)$
to divide the exponential into mean-field and kinetic terms. Since the mean-field is diagonal in real space and the kinetic part is diagonal in momentum space, we can trivially compute the impact of these two exponentials by applying them separately to the real and the momentum space density matrices. By using fast Fourier transforms, we can easily switch from one representation to the other and therefore implement the time evolution in a rather efficient way.

It is worthwhile to comment on the preparation of the initial mean-field density matrix. In order to avoid spurious oscillations in various variables for slabs representing the colliding nuclei, it is convenient to start the time-dependent calculation from a density matrix that is the ground state for the mean-field. To obtain this ground state, we have started from a 1D harmonic oscillator density matrix embedded in a harmonic oscillator potential. Then, following Eq.~\eqref{eq:SOM}, we have very slowly switched off the harmonic oscillator potential, while switching on, at the same time, the mean-field. When this process is carried on adiabatically, the ground state of the self-consistent mean-field density-matrix is reached at the end of the evolution. After this ground state propagator is obtained, we can repeatedly apply the algorithm of Eq.~\eqref{eq:SOM} to the time evolution of the superposition of boosted ground-state density matrices. This in turn gives access to the evolution of one-body observables and total energy.

In Fig.~\ref{fig:denmat}, we show the time evolution for a system where two shells of 1D harmonic oscillators have been filled for each slab.  The slabs have been evolved adiabatically to self-consistency and then boosted by transforming the density matrices with the operators $\text{e}^{\pm i P x}$.  The momentum $P$ corresponded to the center-of-mass energy of $25$ MeV per particle.  It is observed that the two initially separated density distributions eventually overlap and interpenetrate each other. As the system flies apart, a low density region is left behind in which some light fragments seem to be forming. This could be naively associated with some form of multifragmentation process.

One can observe, already in this very crude model, that changing collision energy leads to the appearance of qualitatively different physical processes. At high energies, the slabs interpenetrate each other and break apart, as already commented.  At low collision energies, one can observe the fusion of the slabs as well as the subsequent evolution of the excited state. These results are in qualitative agreement with the findings of the time evolution of 1D slabs within the TDHF formalism in terms of wave-functions \cite{bonche76}.

\begin{figure}
  \includegraphics[width=.33\textwidth]{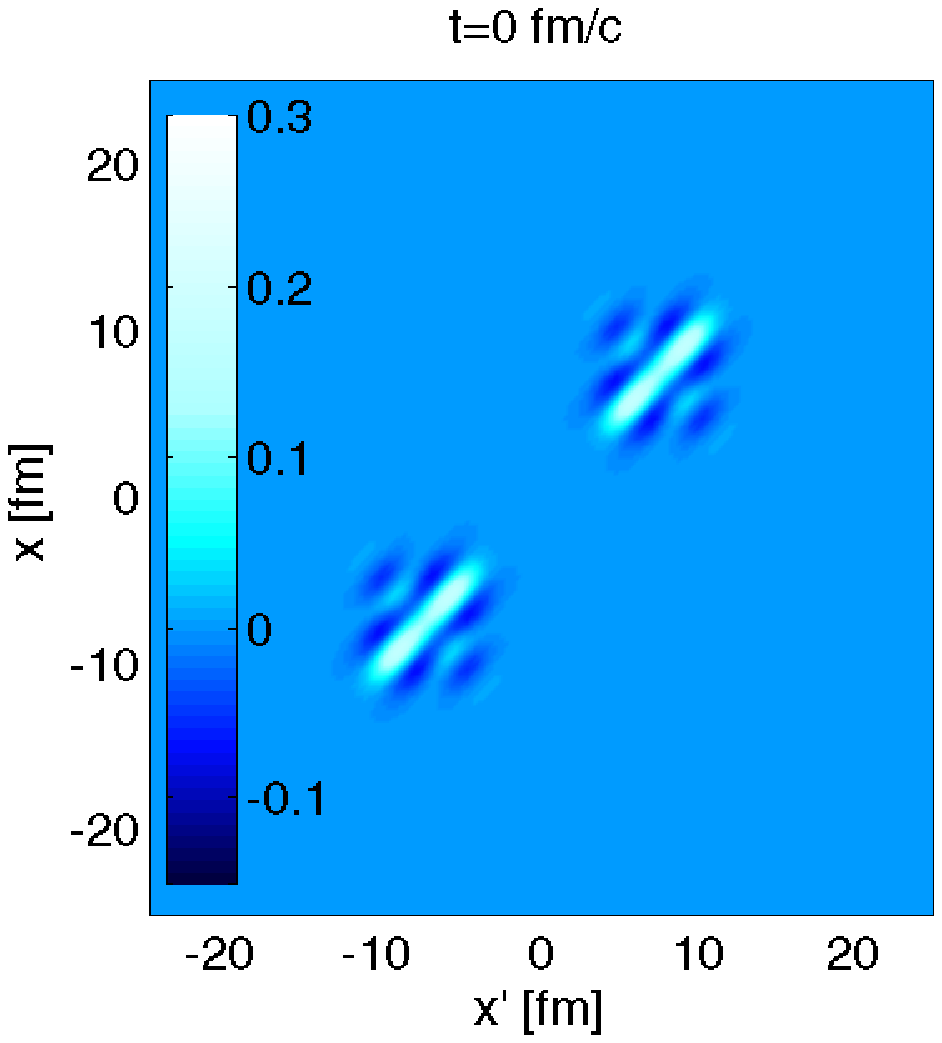}
  \includegraphics[width=.33\textwidth]{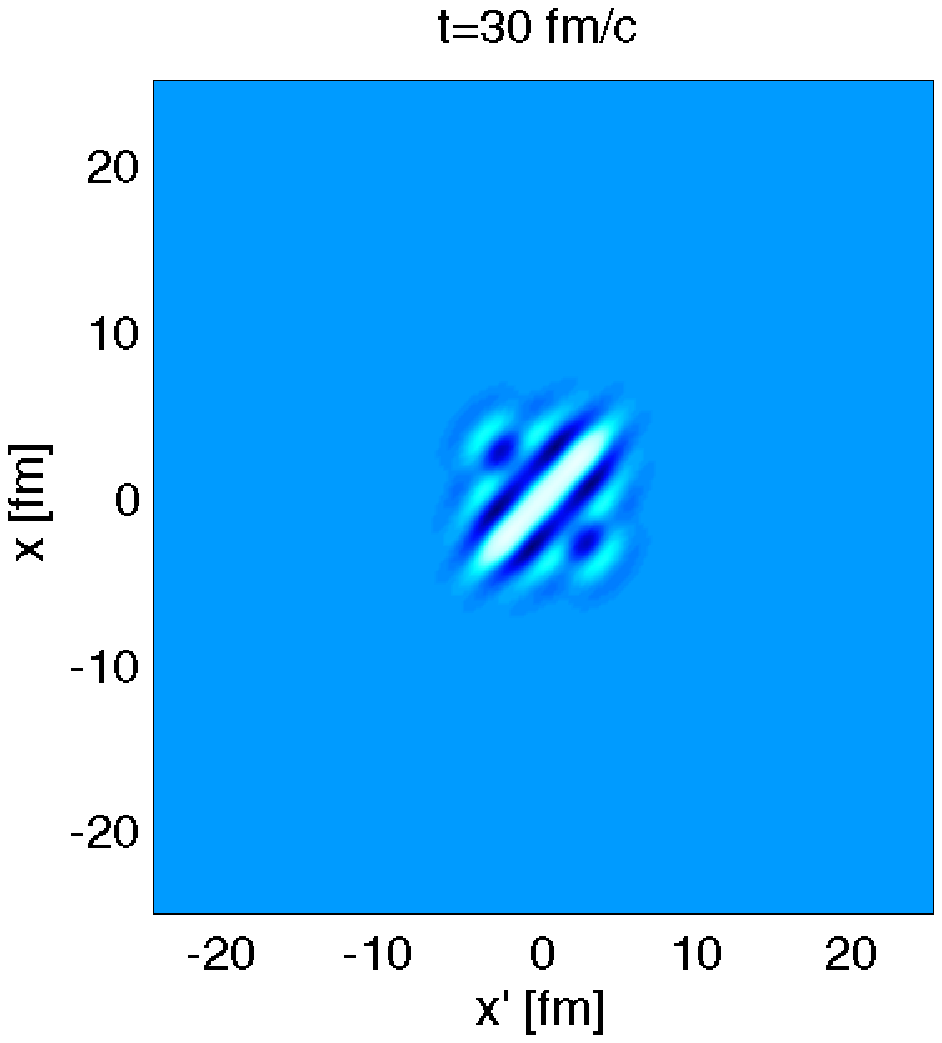}
  \includegraphics[width=.33\textwidth]{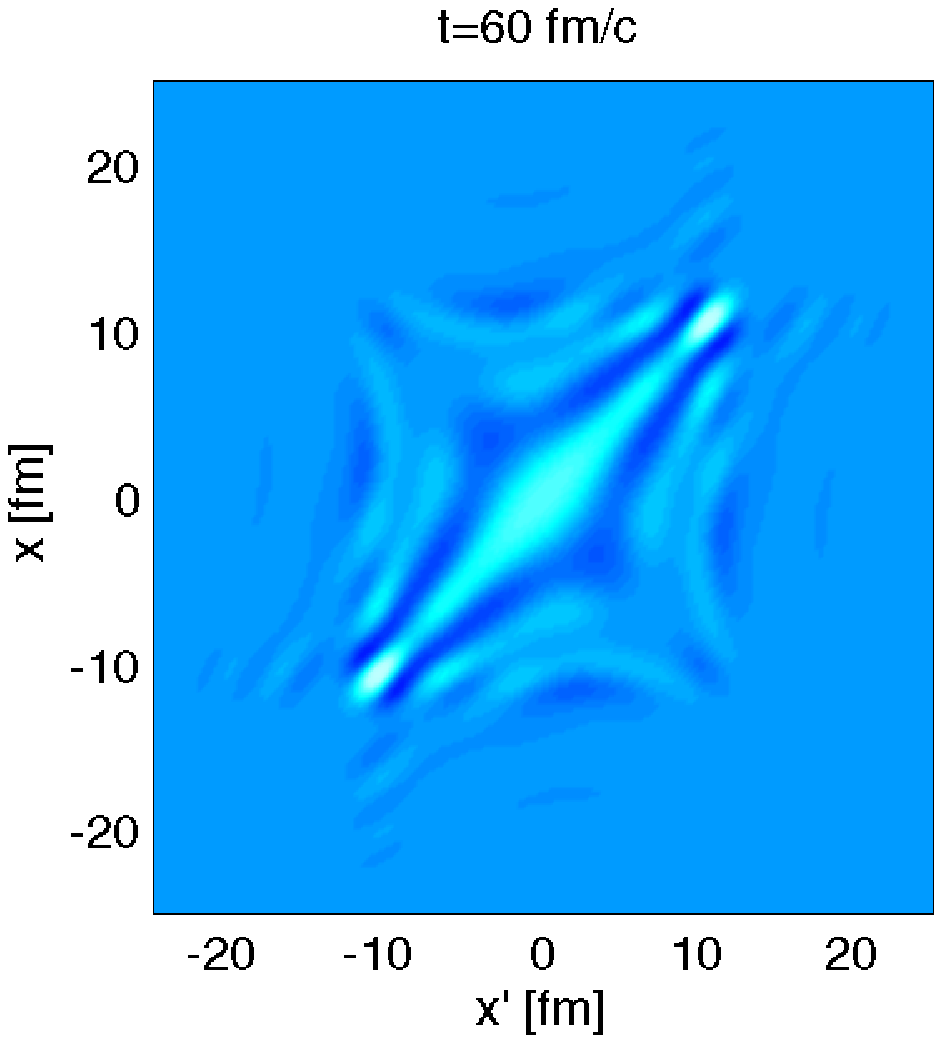}
  \caption{Intensity plots of the real part of $\G^<$ as a function of the $x$ and $x'$ variables. The left, center and right panels represent, respectively, the initial state of the system; the state at $t=30 \, fm/c$ when slabs interpenetrate and the state at $t=60 \, fm/c$ when the slabs move away from each other. Note that the white tones correspond to positive values of the function and the darker ones (dark blue on-line) to negative values.}
  \label{fig:2Dden}
\end{figure}

The time-diagonal density matrix depends on two position variables, $x$ and $x'$. In Fig.~\ref{fig:2Dden}. we show intensity plots for the real part of $\G^<$ at three different stages of the collision, specifically at times $t=0$, $30$ and $60$ fm/c. These plots give an insight on the structure of the Green's functions away from the  $x=x'$ diagonal. Thus, the initial $t=0$ density matrices show some spread in direction perpendicular to the diagonal, an effect which can be understood in terms of the wave-functions in the structure of the matrix.  For one, the support of the matrix is constrained by the support of the wave-functions.  This yields square outlines where the values of the matrix can be significant.  Second, the wave-functions signs interplay, producing a band structure along the diagonal.  In more dimensions, phases would matter.  The changing signs and/or phases are generally associated with the changing momentum content in the wave-functions.  The bigger the momentum spread, the more narrow the spread in $|x - x'|$.  The spread remains similar in the slab collision when the slabs interpenetrate at $t=30 \, \text{fm/c}$.  However, the situation qualitatively changes after the slabs collide and begin to move away from each.  On top of the original structure, the support for the matrix $\G^<$ expands, as illustrated for $t=60 \, \text{fm/c}$.  A larger but fainter square of support emerges, with two corners at the location of slabs on the diagonal and two fainter cross-corners persisting into the future.  While the matrix values still exhibit sharp maxima along the diagonal, the wider region of significant values reflects the fragmentation of original wave-functions. Nucleons from the originally populated states can end up in either of the residues and in the space in-between.  The portions of the wave-functions maintain a phase relationship that would be important for the interference if the portions were ever to recombine.

In the late of the nuclear reactions, though, systems progress through expansions and decays and nearly never recombine.  Thus, following nuclear interpenetration, we expect to be able to neglect the matrix elements far away from the $x=x'$ diagonal without affecting the evolution of the elements close to the axis.  In fact, in our preliminary tests, we find virtually no effect on the evolution of density on the $x=x'$ axis when ignoring elements farther away from the axis than those which are significant for a single slab.  Ignoring the elements amounts to the assumption of decoherence over farther-away distances.

\section{Correlated approximation}

\begin{figure}
  \includegraphics[height=.2\textheight]{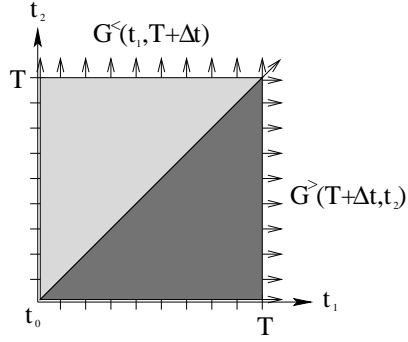}
  \caption{Strategy for solving the Kadanoff-Baym equations in the two time arguments.}
  \label{fig:tevol}
\end{figure}

When integrating the KB equations beyond the mean-field approximation, the SOM method alone is not sufficient.  Even in a 1D case, the solution of the KB equations becomes a relatively involved process.  At present, we are working on the
numerical implementation of the equations for nuclear reactions, following
closely the implementations already existing for uniform matter \cite{danielewicz84b,kohler99} and
generalizing them to the non uniform case.
In this first implementation, we will work within the direct Born approximation for the self-energy.
This is the simplest conserving approximation that includes effects beyond the mean-field.  In the momentum representation, the collisional self-energies are:
\begin{align}
  \Sigma^\lessgtr(p_a,t_a;p_b,t_b) =
\int_{-\infty}^\infty \!\! \frac{\textrm{d} p_1}{2\pi} \int_{-\infty}^\infty \!\! \frac{\textrm{d} p_2}{2\pi}
\, V(p_a-p_1) \, V(p_b-p_2) \, \G^\lessgtr(p_1,t_a;p_2,t_b) \, \Pi^\lessgtr(p_a-p_1,t_a; p_b-p_2, t_b)
\end{align}
where the function $\Pi^\lessgtr$ is a convolution integral of two Green's functions:
\begin{align}
  \Pi^\lessgtr(p_a,t_a;p_b,t_b) =
\int_{-\infty}^\infty \!\!\! \frac{\textrm{d} p_1}{2\pi} \int_{-\infty}^\infty \!\!\! \frac{\textrm{d} p_2}{2\pi} \,
\G^\lessgtr(p_1,t_a;p_2,t_b) \, \G^\gtrless(p_2-p_b,t_b;p_1-p_a,t_a) \, .
\end{align}
In those initial studies, we plan to use a gaussian two-body interaction, $V(p)=V_0 \, \text{e}^{-\frac{\eta^2 p^2}{4} }$,
with parameters
chosen to reproduce semiquantitatively the free-space nucleon-nucleon (NN) scattering cross section.
This should give the first hint on the effect that the inclusion of NN collisions in the formalism has on the time evolution of the different properties of the system.  Since the Born self-energies are given in terms of convolution integrals, they can be efficiently computed using fast-Fourier transform techniques.

To optimally implement the time evolution given by the KB equations, symmetries should be exploited
to minimize some of the computational expenses involved in the calculations. A
particularly useful symmetry is given by the relation:
\begin{align}
[ F^\lessgtr(\mathbf{1}, \mathbf{2}) ]^* = F^\lessgtr(\mathbf{2}, \mathbf{1}) \, ,
\label{eq:sym}
\end{align}
which is fulfilled by both the Green's functions and self-energies. Therefore, one can
store each of the Green's functions in one of the half-planes of the $(t_1,t_2)$
variables and access the other half by using Eq.~\eqref{eq:sym}.
This further suggests the implementation of the time evolution
following the square structure of Fig.~\ref{fig:tevol}. Let us assume that, starting from $t_0$, the Green's functions have been determined for $T \ge t_1, t_2 \ge t_0$ and let us denote the r.h.s.\ of \eqref{eq:kb1} as $I^\lessgtr(\mathbf{1},\mathbf{1}')$.  To the second order in $\Delta t$, the Green's function values at $T+\Delta t$ can be then found from
\begin{align}
\G^>(p_1,T+\Delta t; p_2,t_2) \simeq \text{e}^{i \frac{\varepsilon(p_1) \, \Delta t}{\hbar} } \, \G^>(p_1,T;p_2,t_2)
+ \frac{1}{2} \left[ I^>(p_1,T;p_2,t_2) + I^>(p_1,T+\Delta t; p_2, t_2)  \right]
      \frac{\text{e}^{i \frac{\varepsilon(p_1) \, \Delta t}{\hbar}} - 1 }{\varepsilon(p_1)} \, ,
\label{eq:step}
\end{align}
with $\varepsilon(p_1) = \hbar^2 p^2_1/ 2m$.  An analogous equation holds for
$\G^<$.  A combination of Eqs.~\eqref{eq:kb1} and \eqref{eq:kb1p} yields an algorithm for evolving the functions along the diagonal in Fig.~\ref{fig:tevol}.  Advancing of the functions needs to be repeated to achieve self-consistency as new function values appear on the r.h.s.\ as well, e.g.\ in \eqref{eq:step}.

We expect a general drop in function values \cite{danielewicz84b} for increasing difference $|t_1 - t_2|$.
Similarly, we expect a drop in values for increasing $|x_1 - x_2|$ or $|p_1  - p_2|$, see Fig.~\ref{fig:2Dden}, limiting the size of region within the Green's function matrix, around the diagonal, that needs to be remembered during evolution.  The bookkeeping of values is the particularly costly aspect of the numerical implementation of the
KB equations.

In 1D already, we expect to gain meaningful insights into how the inclusion of correlations makes the dynamics different from the mean-field dynamics or from semiclassical dynamics with nucleon-nucleon collisions.  The inclusion of correlations may cause turn-on problems early on in the evolution and we intend to test the adiabatic switching on of those correlations.  Useful for the 2D and 3D calculations should be an experience with treating the far-off elements of the Green's functions and even the strategies for discretizing the functions.


\begin{theacknowledgments}
Discussions with Dennis Lacroix are gratefully acknowledged.  This work
was partially supported by the National Science Foundation, under
Grant Nos.\ PHY-0555893 and PHY-0551164.
\end{theacknowledgments}



\bibliographystyle{aipproc}   

\bibliography{main}

\hyphenation{Post-Script Sprin-ger}
\begin{thebibliography}{15}
\expandafter\ifx\csname natexlab\endcsname\relax\def\natexlab#1{#1}\fi
\providecommand{\enquote}[1]{``#1''}
\expandafter\ifx\csname url\endcsname\relax
  \def\url#1{\texttt{#1}}\fi
\expandafter\ifx\csname urlprefix\endcsname\relax\def\urlprefix{URL }\fi
\providecommand{\eprint}[2][]{\url{#2}}

\bibitem[Negele(1982)]{tdhf}
J.~W. Negele, \emph{Reviews of Modern Physics} \textbf{54}, 913 (1982).

\bibitem[Tohyama(1987)]{tohyama87}
M.~Tohyama, \emph{Physical Review C} \textbf{36}, 187 (1987).

\bibitem[Lacroix et~al.(2004)]{lacroix04}
D.~Lacroix, S.~Ayik, and P.~Chomaz, \emph{Progress in Particle and Nuclear
  Physics} \textbf{52}, 497 (2004).

\bibitem[Dickhoff and {van Neck}(2005)]{dickhoff05}
W.~Dickhoff, and D.~{van Neck}, \emph{Many-body theory exposed!}, World
  Scientific Publishing, London, 2005.

\bibitem[Kadanoff and Baym(1962)]{kadanoff}
L.~P. Kadanoff, and G.~Baym, \emph{Quantum Statistical Mechanics}, Benjamin,
  New York, 1962.

\bibitem[Baym and Kadanoff(1961)]{baym61}
G.~Baym, and L.~P. Kadanoff, \emph{Physical Review} \textbf{124}, 287 (1961).

\bibitem[Baym(1962)]{baym62}
G.~Baym, \emph{Physical Review} \textbf{127}, 1391 (1962).

\bibitem[Danielewicz(1984{\natexlab{a}})]{danielewicz84a}
P.~Danielewicz, \emph{Annals of Physics} \textbf{152}, 239
  (1984{\natexlab{a}}).

\bibitem[Danielewicz(1984{\natexlab{b}})]{danielewicz84b}
P.~Danielewicz, \emph{Annals of Physics} \textbf{152}, 305
  (1984{\natexlab{b}}).

\bibitem[K\"ohler(1995)]{kohler95}
H.~S. K\"ohler, \emph{Physical Review C} \textbf{51}, 3232 (1995).

\bibitem[Kwong et~al.(1998)]{kwong98}
N.~H. Kwong, M.~Bonitz, R.~Binder, and H.~S. K\"ohler, \emph{Physica Status
  Solidi B} \textbf{206}, 197 (1998).

\bibitem[Dahlen and {van Leeuwen}(2007)]{dahlen07}
N.~E. Dahlen, and R.~{van Leeuwen}, \emph{Physical Review Letters} \textbf{98},
  153004 (2007).

\bibitem[K\"ohler et~al.(1999)]{kohler99}
H.~S. K\"ohler, N.~H. Kwong, and H.~A. Yousif, \emph{Computer Physics
  Communications} \textbf{123}, 123 (1999).

\bibitem[Feit et~al.(1982)]{som}
M.~D. Feit, J.~A. Fleck, and A.~Steiger, \emph{Journal of Computational
  Physics} \textbf{47}, 412 (1982).

\bibitem[Bonche et~al.(1976)]{bonche76}
P.~Bonche, S.~Koonin, and J.~W. Negele, \emph{Physical Review C} \textbf{13},
  1226 (1976).

\end{thebibliography}

\IfFileExists{\jobname.bbl}{}
 {\typeout{}
  \typeout{******************************************}
  \typeout{** Please run "bibtex \jobname" to optain}
  \typeout{** the bibliography and then re-run LaTeX}
  \typeout{** twice to fix the references!}
  \typeout{******************************************}
  \typeout{}
 }

\end{document}